\documentclass{ws-procs9x6}

\begin{document}

\title{Study of leptonic CP violation}

\author{S. NASRI$^A$, J. SCHECHTER$^B$\footnote{\uppercase
{S}peaker} AND S. MOUSSA$^C$}

\address{(A) Department of Physics, \\ 
University of Maryland, \\
College Park, MD 20742-4111, USA.\\
E-mail: snasri@physics.umd.edu}  

\address{(B) Physics Department, \\
Syracuse University, \\
Syracuse, NY 13244-1130, USA. \\
E-mail: schechte@phy.syr.edu}

\address{(C) Department of Mathematics, \\
Faculty of Science, \\
Ain Shams University, Egypt. \\
E-mail: sherif@asunet.shams.eun.eg}

\maketitle

\abstracts{
The ``complementary" Ansatz, Tr($M_\nu$)=0, where $M_\nu$ is
the prediagonal neutrino mass matrix, seems a plausible approximation
for capturing in a self contained way some of the content of
Grand Unification. We study its consequences in the form of
relations between the neutrino masses and CP violation phases.}
 
\section{Introduction}

A favorite topic of discussion at the MRST meetings over many
years has been the prediction of quark and lepton masses and mixings.
Usually, an attempt is made to predict everything at once based
on a suitable guess (Ansatz). Here we discuss a ``complementary"
Ansatz which adds just a little information to the system. This
is \cite{bfns,hz,r} , for the symmetric pre-diagonal neutrino mass matrix, 
$M_\nu$:
\begin{equation}
Tr(M_\nu)=0
\label{ansatz}
\end{equation} 
If, at first, leptonic CP violation is neglected so that $M_\nu$
is diagonalized by a real orthogonal matrix, Eq. (\ref{ansatz})
yields simply
\begin{equation}
m_1+m_2+m_3=0,
\label{masssum}
\end{equation}
where the neutrino masses, $m_i$ can be taken here to be either
positive or negative. 
Now a very recent analysis \cite{mstv} of solar, atmospheric,
reactor and accelerator
neutrino oscillation data (but neglecting LSND) gives the best fit:
\begin{eqnarray}
    m_2^2-m_1^2 &=& 6.9 \times 10^{-5} eV^2, \nonumber \\
    |m_3^2-m_2^2| &=& 2.6 \times 10^{-3} eV^2.
\label{massdifferences}
\end{eqnarray}                                     
Together, Eqs. (\ref{masssum}) and (\ref{massdifferences})
provide three equations for three unknowns. There are two
essentially different types of solutions. Type I is characterized
by $|m_3|$ being largest:
\begin{eqnarray}
m_1=0.0291 \textrm{ eV},\quad  m_2=0.0302 \textrm{ eV},\quad
m_3=-0.0593 \textrm{ eV}, \label{typeone}
\end{eqnarray}
while type II has $|m_3|$ smallest:
\begin{eqnarray}
m_1=0.0503 \textrm{ eV},\quad  m_2=-0.0510 \textrm{ eV},\quad
m_3=0.00068 \textrm{ eV}. \label{typetwo}
\end{eqnarray}                                                          
Here we will, following ref.\cite{nsm} (see this for
further references), discuss the situation when
CP violation effects are included and give an application
to leptogenesis.

\section{Plausibility argument for Ansatz}

SO(10) grand unified theories have the nice feature that they
contain a complete fermion 
generation in a single irreducible representation of the group.
There are three Higgs irreducible representations which can directly
contribute to tree level fermion masses via the Yukawa sector:
the {\bf 10}, the {\bf 120} and the {\bf 126}. In principle any
number of each is allowed. For every Higgs field there is a 3 $\times$3
matrix of unknown coupling constants. The fermion mass matrices 
are linear combinations of these matrices. Clearly a very large number
of different models can be envisioned.

We start from the ``kinematical" relation:
\begin{equation}
Tr(M_{-1/3}-rM_{-1}) \propto Tr(M_{0,LIGHT}),
\label{b-tau}
\end{equation}
which holds when any number of {\bf 10}'s and {\bf 120}'s are
present but only a single {\bf 126}. Here the subscript on
the mass matrix indicates the electric charge of the fermion.
The quantity $r \approx 3$ takes account of running masses
from the GUT scale to about 1 GeV. 
Under the same conditions one also has:
\begin{equation}
 M_{0,HEAVY} \propto M_{0,LIGHT}.
\label{clone}
\end{equation}
Note that the physical light neutrino mass matrix, $M_\nu$
is given by the well known formula:
\begin{equation}
M_\nu \approx M_{0,LIGHT} - M_{DIRAC}^T M_{0,HEAVY}^{-1} M_{DIRAC}.
\label{seesaw}
\end{equation}
The initial assumption we shall make is that the second, ``see-saw"
term in Eq. (\ref{seesaw}) is small compared to $M_{0,LIGHT}$.

Now, it has been known for a long time that the quark mixing
matrix is of the form $diag(1,1,1) + O(\epsilon)$. Thus
it was very surprising when analysis of neutrino oscillation
observations showed that the lepton mixing matrix is not at all
close to the unit matrix but rather has large (12) and (23)
mixing elements. In a GUT framework, this suggests that a first
approximation to the prediagonal mass matrices might be to take the
charged fermion mass matrices to be diagonal while the neutrino 
mass matrix would presumably differ drastically from the diagonal
form. As examples we would set $M_{-1} \approx diag(m_e, m_\mu, m_\tau)$
and $M_{-1/3} \approx diag(m_d, m_s, m_b)$. Substituting these into
the left hand side of Eq. (\ref{b-tau}) shows it to be about $(m_b-3m_\tau)$,
which is about zero. Hence the right hand side should also be about zero
as should $Tr(M_\nu)$ in the non-seesaw dominance case. Although this
is clearly an approximation, it seems likely to be close to the
physical situation in the same sense as $m_b \approx 3m_\tau$. Of course,
the approximation gets better as the mixing matrices
needed to bi-diagonalize the charged lepton mass  matrix
get closer to the unit matrix. For simplicity, in what follows  
we shall also approximate these matrices to equal
the unit matrix.

\section{Parameterized Ansatz equation}
The prediagonal neutrino mass matrix may be brought to diagonal
form by a transformation:
\begin{equation}
U^{T}M_{\nu} U = {\hat M_\nu} = diag (m_1, m_2, m_3),
\label{transf}
\end{equation}
where $U$ is a unitary matrix. $M_\nu$ is a symmetric but
complex matrix which has in general 12 real parameters. This
equals the sum of the three parameters from $m_i$ and the
nine parameters from $U$. Now the observable lepton mixing
matrix, $K$ is given \cite{bfns} by $K=\Omega^{\dagger}U$,
where we have just agreed to approximate the charged lepton
diagonalizing matrix factor, $\Omega$ to be
essentially the unit matrix (or alternatively we could choose
to work in a basis where the charged leptons are diagonal).
Thus we replace U by the observable matrix $K$. $K$ is
parameterized in a conventional way as $K=K_{exp}\omega^{-1}_0(\tau)$,
where $\omega_0(\tau)=diag(e^{i\tau_1},e^{i\tau_2},e^{i\tau_3})$
with $\tau_1+\tau_2+\tau_3=0$ and: 
\begin{equation}
K_{exp}=\left[ \begin{array}{c c c}
c_{12}c_{13}&s_{12}c_{13}&s_{13}e^{-i\delta}\\
-s_{12}c_{23}-c_{12}s_{13}s_{23}e^{i\delta}&c_{12}c_{23}-s_{12}s_{13}s_{23}
e^{i\delta}&c_{13}s_{23}\\
s_{12}s_{23}-c_{12}s_{13}c_{23}e^{i\delta}&-c_{12}s_{23}-s_{12}s_{13}c_{23}
e^{i\delta}&c_{13}c_{23}\\
\end{array} \right]
\label{usualconvention}
\end{equation}
where $s_{ij} = sin \theta_{ij}\; and \; c_{ij} = cos
\theta_{ij}$. As written, the matrix $K$ is parameterized by three
angles and three independent CP violating phases. To get the most general
$K$ three more phases are required; these can be inserted for example
 \cite{bfns} by multiplying $K$ on the left by $\omega_0(\sigma)$.
However these phases can always be canceled by rephasing the (diagonal)
charged lepton fields which sit to the left of $K$. Thus even if the
phases, $\sigma_i$ were included in $U$ there would always be 
an allowed choice of charged lepton phases which would cancel their
effect when we get restrictions (as we shall) on the physical $K$. 
 Taking this into account the Ansatz reads in terms of physical
quantities:            
\begin{equation}
Tr({\hat M_\nu}K_{exp}^{-1}K_{exp}^*\omega_0(2\tau))=0.
\label{ansatzwithKexp}
\end{equation}                           
In more detail it is:
\begin{eqnarray}
m_1e^{2i\tau_1} \left[ 1 -2i(c_{12}s_{13})^2sin{\delta}e^{-i\delta} 
\right]
 +\nonumber\\
m_2e^{2i\tau_2} \left[ 1 -2i(s_{12}s_{13})^2sin{\delta}e^{-i\delta} 
\right]
\nonumber +\\
m_3e^{2i\tau_3} \left[ 1 +2i(s_{13})^2sin{\delta}e^{i\delta} \right] = 0.
\label{paramansatz}
\end{eqnarray}
In this equation we can choose the diagonal masses $m_1, m_2, m_3$ to be
real positive. The mixing angles are known 
from the best fit \cite{mstv},                                                              
\begin{equation}
s_{12}^2 = 0.30,\quad  s_{23}^2 = 0.50,\quad  s_{13}^2 = 0.003,
\label{exptmixingangles}
\end{equation}                               
wherein the first two have about 25 per cent uncertainty
while the third is just known to be small.

Together Eqs. (\ref{massdifferences}) and (\ref{paramansatz}) are now seen 
to provide
4 real equations for the 6 unknowns: $m_1,m_2,m_3,\delta,\tau_1,\tau_2$.
Thus further assumptions are needed to make some predictions.
We already saw that assuming all the CP phases to vanish gives three
equations for three unknowns. If we assume only the ``conventional"
CP phase $\delta$ not to vanish there are four equations for four
unknowns, with the results described in \cite{nsm}. This case
would become trivial in the limit where $s_{13}^2$ is assumed to vanish.
There is no reason to expect it to vanish exactly but,
considering our lack of knowledge, that seems to be also an interesting
assumption to investigate. According to Eq. (\ref{paramansatz}) it yields
the same result as setting $\delta=0$. Then we have 4 equations
for 5 unknowns and can get a one parameter family of solutions.
 
\section{Family of Majorana phases}

In this case, Eq. (\ref{paramansatz}) takes the form
\begin{equation}
m_1e^{2i\tau_1}+m_2e^{2i\tau_2}+m_3e^{2i\tau_3}=0,
\label{majtypeviol}
\end{equation}
which corresponds, as illustrated in Fig. 1, to a triangle in the complex 
plane.

\begin{figure}[htbp]
%\begin{center} 
%\epsfxsize = 12cm
%\ \epsfbox{test1b.eps}
%\mbox{\epsfig{file=test1b.eps,height=4in,width=4in,angle=0}}
%\end{center}
\centering
{\includegraphics[width=6.00cm,height=5.50cm,clip=true]{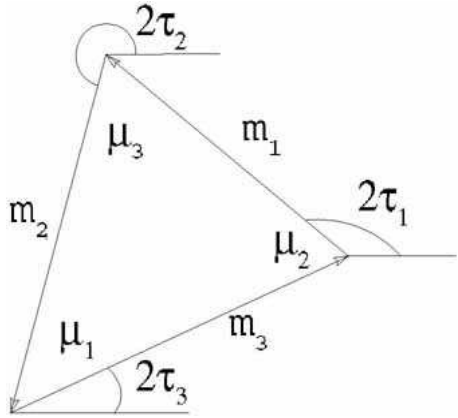}}
\caption[]{Vector triangle representing Eq. (\ref{majtypeviol}).}
\label{fig:1}   
\end{figure}

We proceed to obtain the family of solutions by assuming a value for
$m_3$, getting $m_1$ and $m_2$ from Eq. (\ref{massdifferences})
and finally by solving for the two interior angles $\mu_1$ and $\mu_2$
using trigonometry. Interesting CP violation quantities turn out to be:
\begin{eqnarray}
sin[2(\tau_1-\tau_2)]&=&-sin(\mu_1+\mu_2), \nonumber\\
sin[2(\tau_1-\tau_3)]&=&sin\mu_2, \nonumber\\
sin[2(\tau_2-\tau_3)]&=&-sin\mu_1,
\label{usefulsines}
\end{eqnarray}

The criterion for the existence of CP violation effects is the
area of the triangle being different from zero. We may express
the area as:
\begin{equation}
Area=\frac{1}{4}\left([(m_1+m_2)^2-m_3^2][m_3^2-(m_1-m_2)^2]\right)^{1/2}.
\label{secondarea}
\end{equation}
Now we see that the vanishing of the first factor corresponds to
the type I real solution while the vanishing of the second factor
corresponds to the type II real solution. Furthermore, for a solution to
exist, the argument of the square root should be positive.
With the second factor, that establishes the minimum allowed value
of $m_3$ while with the first factor, that establishes the minimum
value of $m_3$ which allows a type I solution.
    
The table below shows a ``panorama" of solutions decreasing from $m_3 
=0.3$ eV,
(which is about the highest value compatible with the cosmology bound
\cite{cosmobound} that the sum of the neutrino masses be
 less than about 1 eV) to the lowest value imposed by the model.
 In the type I solutions
 $m_3$ is the largest mass
while in the type II solutions $m_3$ is the smallest mass. For each
 value of $m_3$, the values of the model predictions for $m_1$
 and $m_2$ as well as the
 internal angles $\mu_1$ and $\mu_2$ are given. The model
prediction for the  neutrinoless double beta decay quantity $|m_{ee}|$
is next shown. Finally, the last column shows the estimated
 lepton asymmetries due to the decays of the heavy
 neutrinos. Note that the reversed sign
of lepton asymmetry is also possible.                     
           
\begin{table}[htbp]
%\tbl{Panorama of solutions as $m_3$ is lowered from about the
%highest value which is experimentally reasonable to the
%lowest value imposed by the model. In the type I solutions
% $m_3$ is the largest mass
%while in the type II solutions $m_3$ is the smallest mass. For each
% value of $m_3$, the values of the model predictions for $m_1$
% and $m_2$ as well as the
% internal angles $\mu_1$ and $\mu_2$ are given. The model
%prediction for the  neutrinoless double beta decay quantity $|m_{ee}|$
%is next shown. Finally, the last column shows the estimated
% lepton asymmetries due to the decays of the heavy
% neutrinos. Note that the reversed sign
%of lepton asymmetry is also possible.}
\begin{center}                                       
\begin{tabular}{lllll}
\hline \hline  & $m_1,m_2,m_3$ in \textrm{ eV} & $\mu_1,\mu_2$ 
rad. & $|m_{ee}|$ &
  $\epsilon_1$, $\epsilon_2$, $\epsilon_3$ \\
\hline
\hline
I & .2955, .2956, .300 & 1.038, 1.039 & .185 &  .342, .433, .017
\\
II & .3042, .3043, .300 & 1.055,1.056 & .187 & .330, .426, -.0172
\\
I & .0856, .0860, .100 & 0.946, 0.952 & .058 & .138, .060, .00137
\\
II & .1119, .1122, .100 & 1.106, 1.111 & .065 & .194, .088,-.0024
\\
I & .0305, .0316, .060 & 0.258, 0.268 & .030 &  .00982, .00422,
.00004 \\
II & .0783, .0787, .060 & 1.172, 1.187 & .043 & .094, .041,-.0011
\\
I & .0291, .0302, & .000552,  & .030 & 
1.96 $\times 10^{-6}$,\\
 &  0.0592715649   &   0.000574   & &.84 $\times 10^{-6}$, .71 $\times 
10^{-7}$  \\
II & .0774, .0782,  & 1.174, 1.188 & .042 &.047, .020,
-.0011  \\
 & 0.0592715649 &  & & \\
II & .0643, .0648, .040 & 1.243, 1.268 & .033 & .052,
.023,-.000681  \\
II & .0541, .0548, .020 & 1.355. 1.442 & .024 & .018,
.0078,-.000335  \\
II & .0506, .0512, .005 & 1.386, 1.658 & .021 &
 .0057, .0025,-.0000824  \\
II & .0503, .0510, .001 & 0.814, 2.313 & .021 & .00073,
.00031, \\
 & & & &-.0000122 \\
II & 0.0503, 0.0510, & 0.051361,  & .021 &
 .0000348, .0000150,   \\
 &  0.0006996 & 3.089536 & &-0.601 $\times 10^{-6}$ \\
\hline
\hline
\end{tabular}
\end{center}
%\caption[]{Panorama of solutions as $m_3$ is lowered from about the
%highest value which is experimentally reasonable to the
%lowest value imposed by the model. In the type I solutions
% $m_3$ is the largest mass
%while in the type II solutions $m_3$ is the smallest mass. For each
% value of $m_3$, the values of the model predictions for $m_1$
% and $m_2$ as well as the
% internal angles $\mu_1$ and $\mu_2$ are given. The model
%prediction for the  neutrinoless double beta decay quantity $|m_{ee}|$
%is next shown. Finally, the last column shows the estimated  
% lepton asymmetries due to the decays of the heavy
% neutrinos. Note that the reversed sign
%of lepton asymmetry is also possible.}
\label{trianglesampler}
\end{table}

Notice that when $m_3$ is decreased to about 0.0593 eV,
 we get to the real type I case (no CP violation). Below this
value of $m_3$ only the type II solutions exist. At $m_3$ 
about $7 \times 10^{-4}$ eV, we get the real type II case
and no solutions exist for $m_3$ below this value. In the $m_3$
regions just above the two real cases we can evidently tune the
CP violation phases continuously to be as small as desired.

An interesting application of the model is to  neutrinoless
double beta decay (for example, the decay ${}^{76}Ge \rightarrow
{}^{76}Se +e^-+e^-$). 
 The current experimental bound on the amplitude factor \cite{ndbdexpt}
is: $|m_{ee}| < (0.35 \rightarrow 1.30)$ eV, 
where
\begin{equation}
|m_{ee}| = \sum 
|m_i(K_{exp1i})^2e^{-2i\tau_i}|.
\label{ndbd2}
\end{equation}                                                            
From the table we see that the predicted values of $|m_{ee}|$
are typically about one order of magnitude below the experimental
bound. Furthermore, the predicted values do not vary drastically
for $m_3$ less than about 0.1 eV.

\section{Estimate for leptogenesis}
    An intriguing possibility for learning more about leptonic
CP violation is the study of the proposed leptogenesis mechanism
for generation of the present baryon number asymmetry of the universe.
According to this scheme, the lepton number violating decays of the
 heavy neutrinos at a high temperature (early universe) establish
a lepton asymmetry which gets converted as the universe cools,
 through a (B+L) violating but (B-L) conserving ``sphaleron"
interaction to a baryon asymmetry. References and a rough estimate
in the present framework are given in ref. \cite{nsm}. According
to Eq. (\ref{clone}) the heavy neutrino masses are supposed to
be proportional to the light ones here and have the same diagonalizing 
matrix, $U$. The effective
term for calculating the heavy neutrino decays at very high temperature
is 
\begin{equation}
{\it L}_{YUKAWA} = -\sum {\bar L}_ih_{ij}\Phi^c{\hat N}_j + H.c.,
\label{decayvertex}
\end{equation}
where
\begin{equation}
h_{ij} \approx \frac{M_{2/3i}K_{expij}e^{-i\tau_j}}{<\phi^0>r^{\prime}}.
\label{hmatrix}
\end{equation}
The quantities needed for the calculation are the matrix
products $(h^{\dagger}h)_{ij}$; it is thus seen that the effect of a diagonal
matrix of phases multiplying $K_{expij}$ on the left would cancel out.  
   The lepton  CP asymmetry $\epsilon_i$, due to the decay of the i{\emph th}
heavy neutrino, is defined as the ratio of decay widths:
\begin{equation}
\epsilon_i = \frac{\Gamma(N_i\rightarrow L+\Phi)-\Gamma(N_i \rightarrow
{\bar L}+{\bar \Phi})}{\Gamma(N_i\rightarrow L+\Phi)+\Gamma(N_i 
\rightarrow
{\bar L}+{\bar \Phi})}.
\label{defineepsilon}
\end{equation}
In this equation $L+\Phi$ stands for all lepton- Higgs pairs
 of the types $e^-_j +\phi^+$
and $\nu_j + {\bar \phi}^0$. This is an effect which violates C and CP
conservation, in agreement with the requirement of Sakharov.
 The numerical values of the $\epsilon_i$, which 
depend on the ratios of heavy neutrino masses rather than their
absolute values, are displayed in the last column of the table.
Notice that Eq. (\ref{decayvertex}) represents the same term which 
generates the Dirac mass, $M_{DIRAC}$ in Eq. (\ref{seesaw}).
Since our motivation for the Ansatz assumes dominance of the
non seesaw term, this feature requires \cite{nsm} the heavy 
neutrino mass scale to be suitably large. This scale plays a role
in the estimation of the present baryon to photon ratio, $\eta_B$
of the universe which is obtained by convoluting the $\epsilon_i$
with factors obtained by solving the Boltzmann evolution equations
for the (B-L) asymmetry. It turns out \cite{nsm} that for typical values
of the parameter, $m_3$ in the table, $\eta_B$ is considerably
larger than its experimental value \cite{cosmobound}
 of about $6.5 \times 10^{-10}$.
Thus agreement with experiment requires tuning close to the two
real type solutions; the correct order of magnitude is obtained
when either $m_3 \approx$ 0.059 eV (type I) or $m_3 \approx$ 0.005 eV
 (type II).
                                                                             
We thank D. Black, A. H.
Fariborz, C. Macesanu , M. Trodden and D. Schechter
 for their help. 
 The work of S.N is supported by National Science foundation grant No. 
PHY-0099544.
The work of J.S. is supported in part by the U. S. DOE under
Contract no. DE-FG-02-85ER 40231.
                                                                               
Finally, we are pleased to take this opportunity to express our wishes
for Happy Birthday, Good Health and Continued Important Contributions 
to Physics
to PAT O'DONNELL and to HARRY LAM.


\begin{thebibliography}{0}

\bibitem{bfns}D. Black, A. H. Fariborz, S. Nasri and J. Schechter,
Phys. Rev. {\bf D62}, 073015 (2000).   

\bibitem{hz}X.-G. He and A. Zee, Phys. Rev. {\bf D68}, 037302, (2003). 

\bibitem{r}W. Rodejohann, Phys. Lett. {\bf B579}, 127 (2004). 

\bibitem{mstv} M. Maltoni, T. Schwetz, M. A. Tortola and J. W. F. Valle,
arXiv:hep-ph/0309130. 

\bibitem{nsm}S. Nasri, J. Schechter and S. Moussa,
arXiv:hep-ph/0402176.

\bibitem{cosmobound}D. N. Spergel et al, arXiv:astro-ph/0302209;
S. Hannestad, arXiv:astro-ph/0303076.

\bibitem{ndbdexpt}H. V. Klapdor-Kleingrothaus et al, Eur. Phys. J. {\bf 
A12}, 147 (2001). 


\end{thebibliography}
\end{document}